\documentclass[aps,prb,twocolumn,groupedaddress]{revtex4}

\usepackage{graphicx}
\newcommand{\tabcharac}{
\begin{table*}[t]
\begin{center}
\caption{Characteristic parameters of the melt-textured bulk samples (the first five) and
of the c-axis oriented films (the last two) of YBCO studied here: $t$, $w$ and $l$ are
respectively the thickness, width and length of the samples. $T_{c0}$ is the offset
temperature where the transition to the normal states begins. $\rho_n$ is the average
resistivity extrapolated from the normal-state resistivity over the temperatures where
$E$--$J$ curves have been measured. $J_c$ is the critical current density obtained by a
10 $\mu$V/cm threshold criterion and $J^*$ is the quenching current density determined
from a deviation criterion from a background functional expression (see text for
details).}
\smallskip
\label{SamCharTable}
\begin{tabular}{lccccccc}
\hline
 Sample & $t$ & $w$ & $l$ & $T_{c0}$ & $\rho_n$ & $J_{c}(0.95\;T_{c0})$ & $J^*(0.95\;T_{c0})$\\
  & {\footnotesize (mm)} & {\footnotesize (mm)} & {\footnotesize (mm)} & {\footnotesize (K)} & {\footnotesize ($\mu\Omega$cm)} & {\footnotesize (A/cm$^2$)} & {\footnotesize (A/cm$^2$)} \\
 \hline
 A21 & 0.65 & 0.80 & 8.1 & 90.4 & 112 & 0.7 $10^4$ & 3.9 $10^4$ \\
 B31 & 0.20 & 0.75 & 7.7 & 89.4 & 78 & 1.5 $10^4$ & 4.0 $10^4$ \\
 B44 & 0.35 & 0.65 & 11.2 & 88.9 & 90 & 1.6 $10^4$ & 4.9 $10^4$ \\
 C11 & 0.40 & 0.60 & 5.7 & 88.7 & 100 & 1.7 $10^4$ & 8.2 $10^4$ \\
 C22 & 0.35 & 0.40 & 7.0 & 88.4 & 43 & 4.2 $10^4$ & 13.4 $10^4$\\
 Sys116 & 1.5 10$^{-4}$ & 10$^{-2}$ & 5 10$^{-2}$ & 90.2 & 102 & 5.0 $10^5$ & 1.3 $10^6$ \\
 Sy3 & 1.9 10$^{-4}$ & 10$^{-2}$ & 1.0 & 91.0 & 53 & 5.2 $10^5$ & 1.5 $10^6$\\
 \hline
\end{tabular}
\end{center}
\end{table*}}

\newcommand{\figVt}{
\begin{figure}
\includegraphics[width=7.3cm]{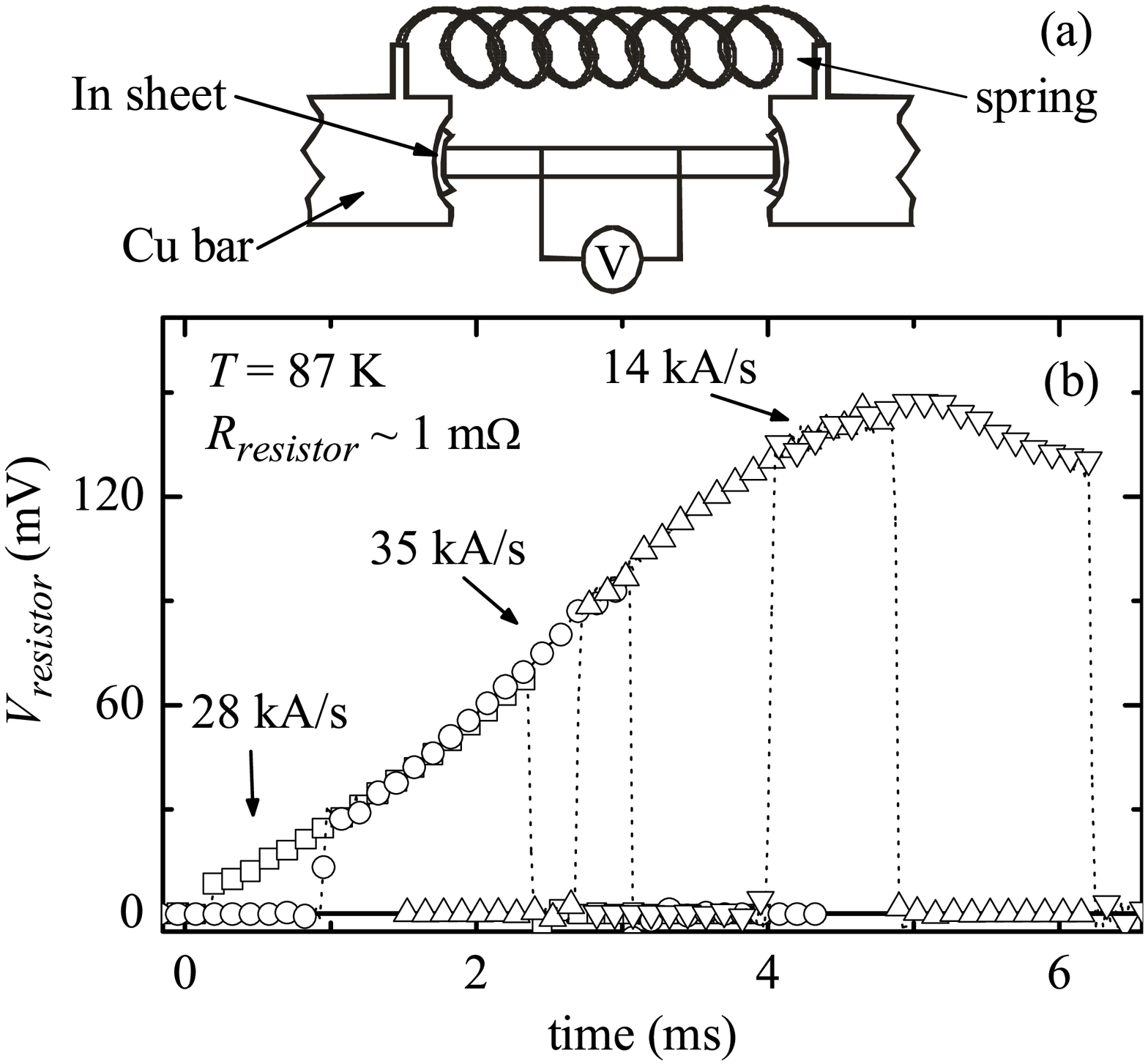}
\caption{\label{figPulseShape} (a) Scheme of the experimental setup for melt-textured
samples. (b)Voltage in the reference resistor versus time for different pulses applied to
one of the melt-textured samples. The current raising rate is determined by the power
supply properties and depends on the total resistance of the circuit. The current
decrease in the last pulse is due to a high increase in the sample resistance as
transition to the normal state occurs.}
\end{figure}}

\newcommand{\figInduct}{
\begin{figure}
\includegraphics[width=7.3cm]{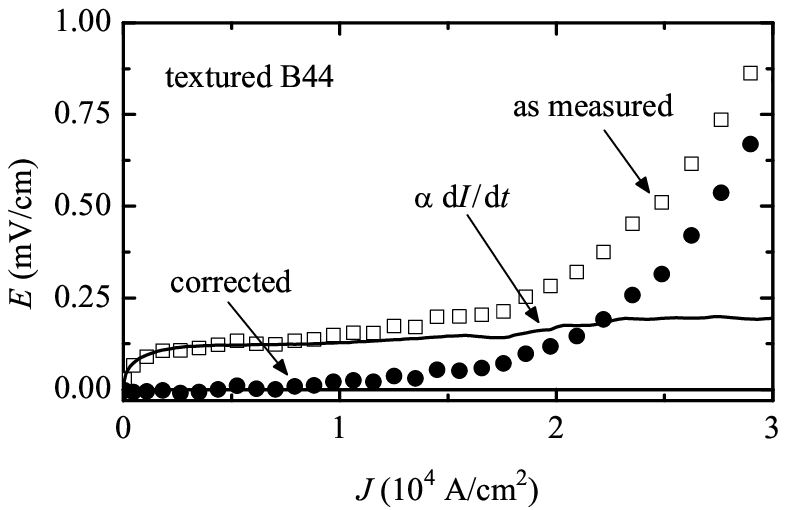}
\caption{\label{figEindvsJ}  Inductive electric field correction for the sample B44 at
85.5 K. d$I$/d$t$ is the numeric derivative of current. Inductance is deduced from the
data at low current density, where the electric field is purely inductive. For this
sample, the inductance is 1.3 nH.}
\end{figure}}

\newcommand{\figSpets}{
\begin{figure}
\includegraphics[width=7.3cm]{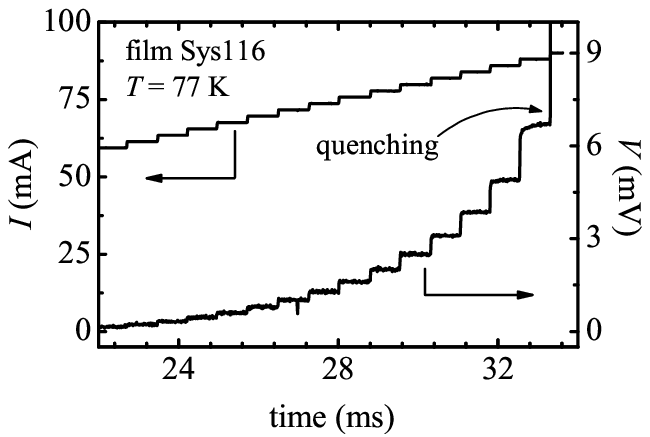}
\caption{\label{figSteps}  Evolution of current and voltage with time for the film
Sys116, during one of the stepped ramps applied in the high dissipation regime. Both
quantities are monitored during the ramp with a high speed DAQ. Voltage is stabilized in
less than 1~ms, indicating that a quasi steady state is reached, though less so, very
near the quenching point.}
\end{figure}}

\newcommand{\figEJTte}{
\begin{figure}
\includegraphics[width=7.3cm]{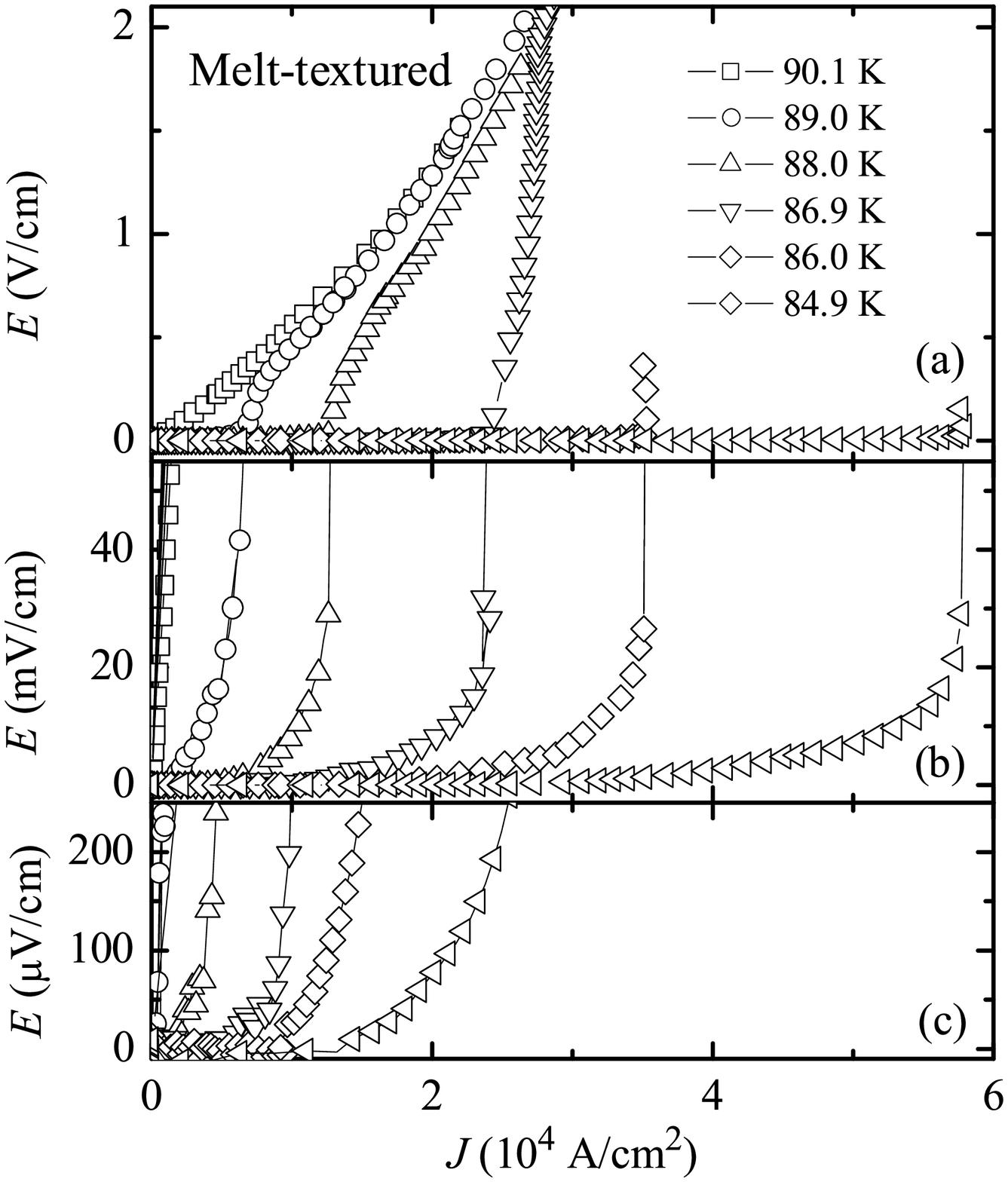}
\caption{\label{figEJTText} CVCs of melt-textured sample B31 for different temperatures,
at three scales: high (a), medium (b) and low (c) voltage, highlighting, respectively,
the abrupt transition to the nearly normal state, non linear regime and zero resistance.}
\end{figure}}

\newcommand{\figEJTfi}{
\begin{figure}
\includegraphics[width=7.3cm]{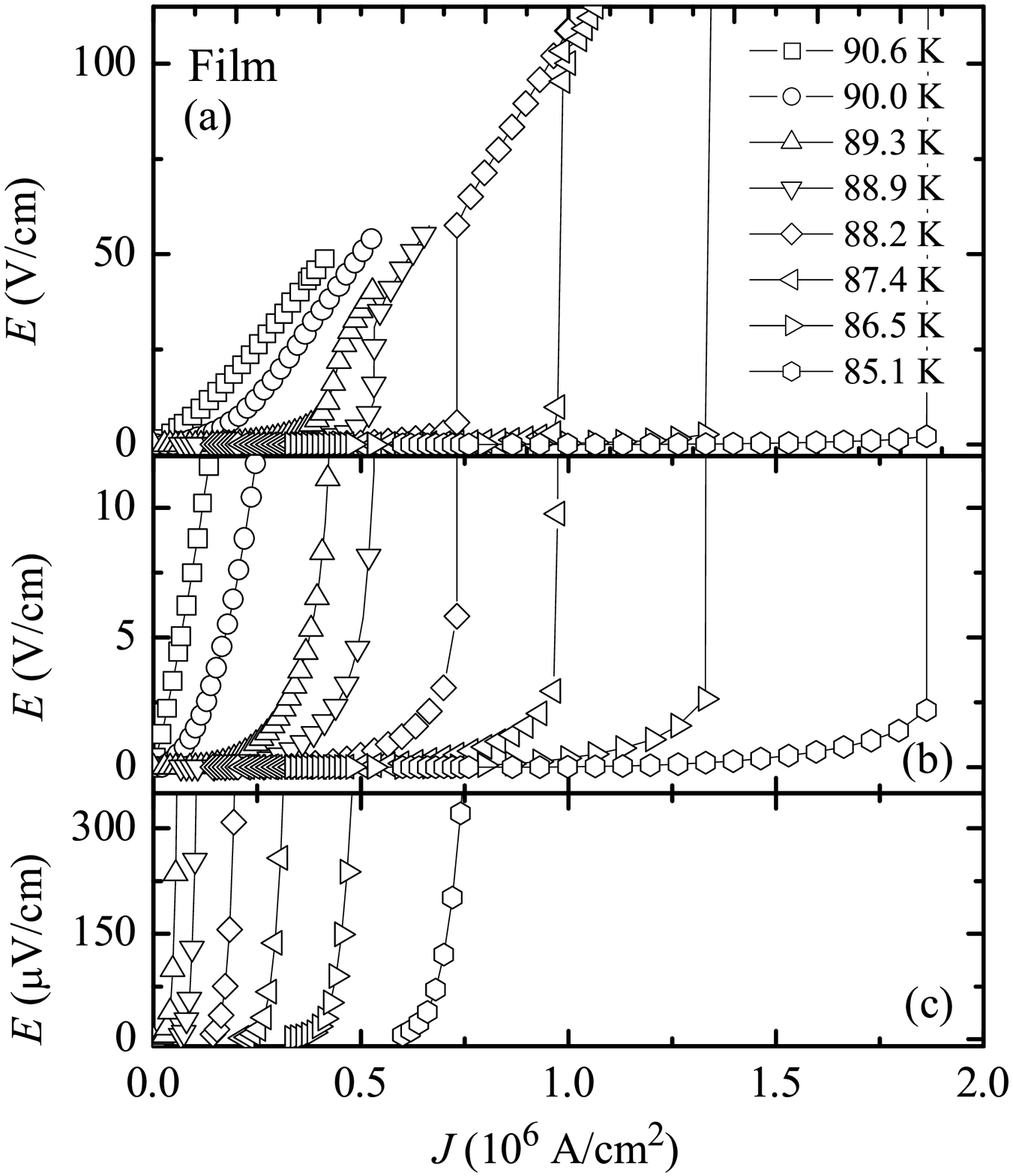}
\caption{\label{figEJTFilm} CVCs at different temperatures for the thin film Sys116.
Again the three dissipation regimes: the abrupt transition to a nearly normal state,
non-linear regime and non zero resistance are stood out in parts (a), (b) and (c).}
\end{figure}}

\newcommand{\figJcvsT}{
\begin{figure}
\includegraphics[width=7.3cm]{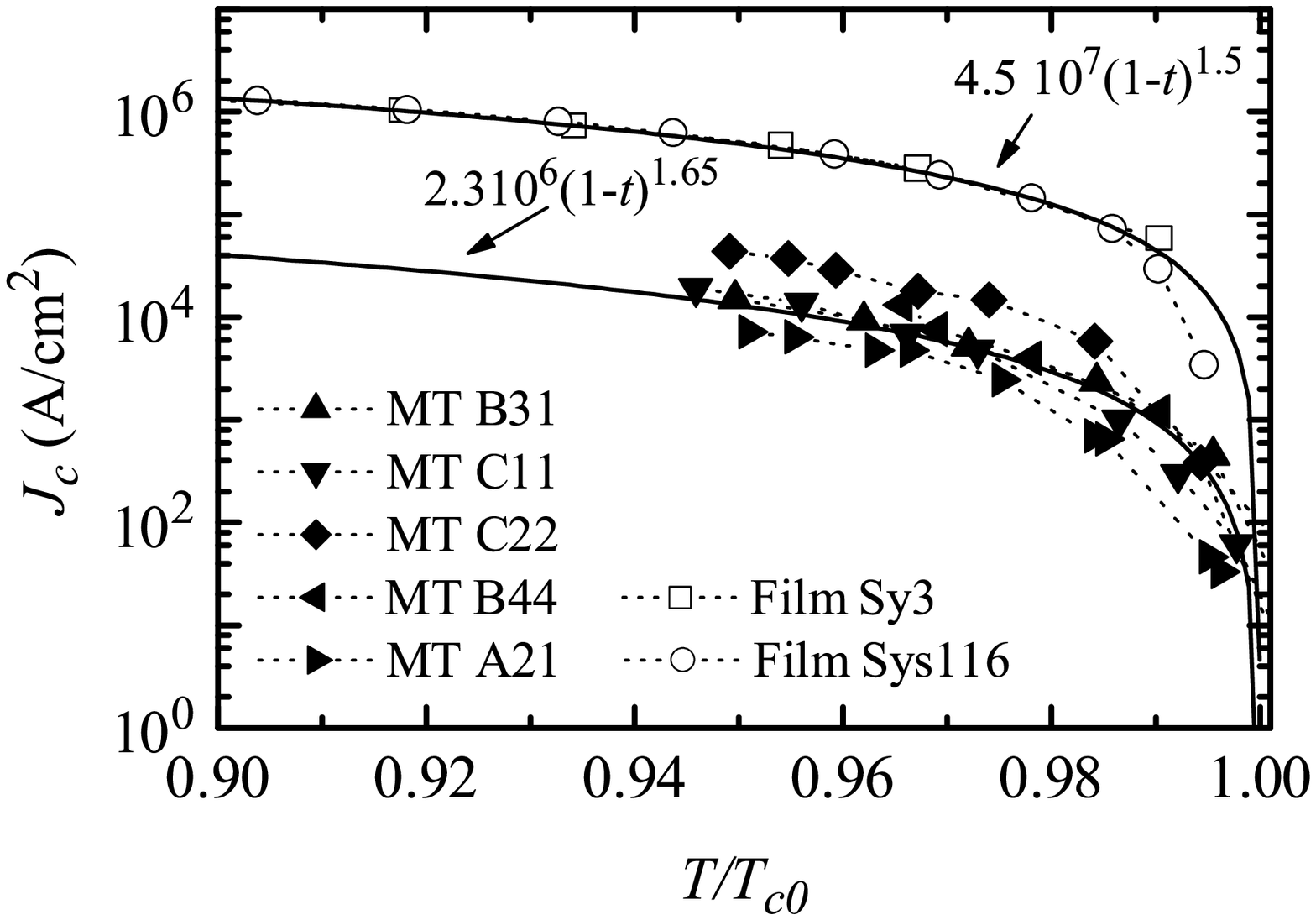}
\caption{\label{figJcvsT} Critical current density, determined by a threshold criterion
of 10 $\mu$Vcm, vs temperature for all the studied samples. Solid lines are the best fit
of the functional form $J_{c0}(1-t)^m$ to the group of results for bulks and films.
Dotted lines are guides for the eye.}
\end{figure}}

\newcommand{\figFitOneT}{
\begin{figure}
\includegraphics[width=7.3cm]{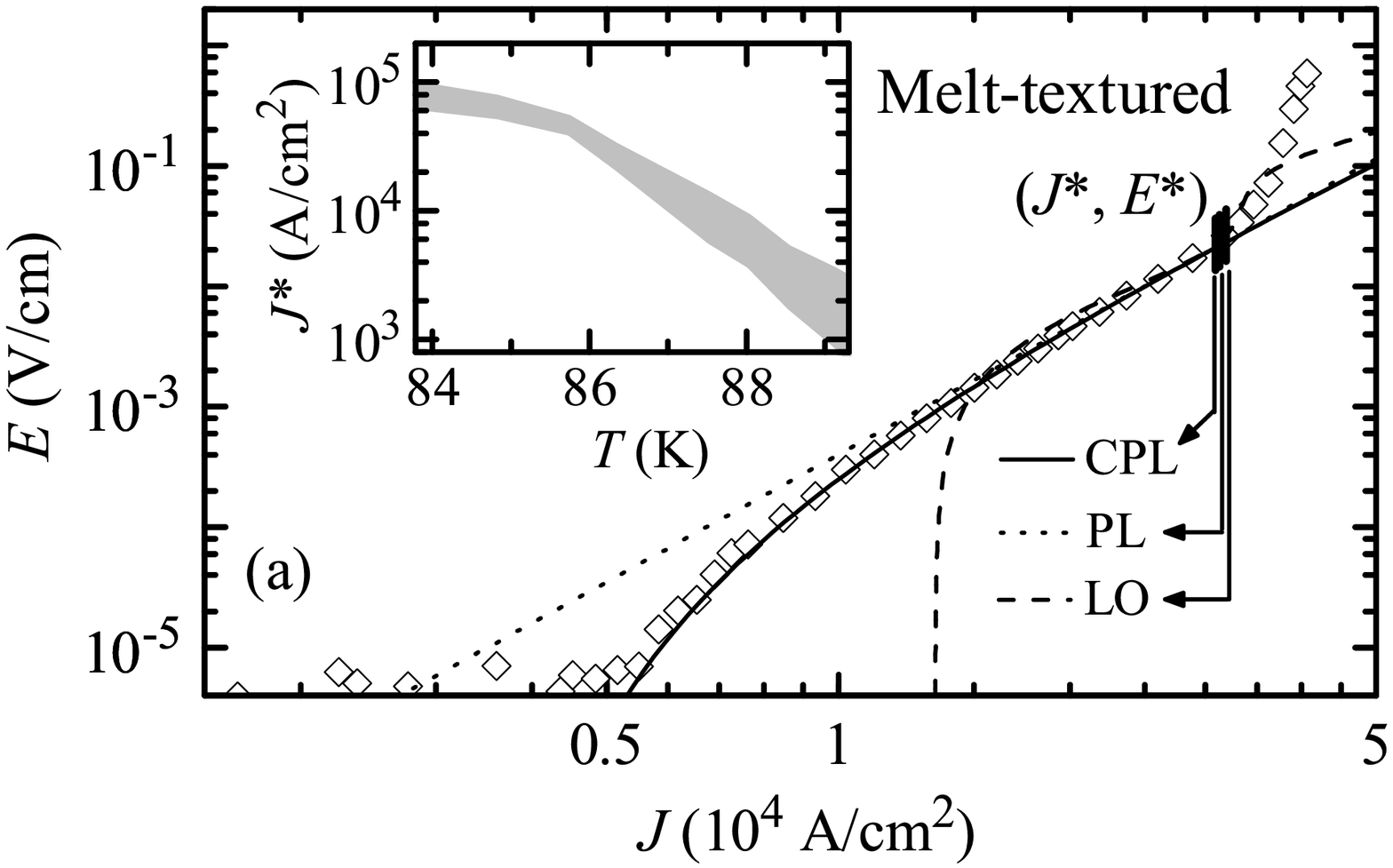}
\includegraphics[width=7.3cm]{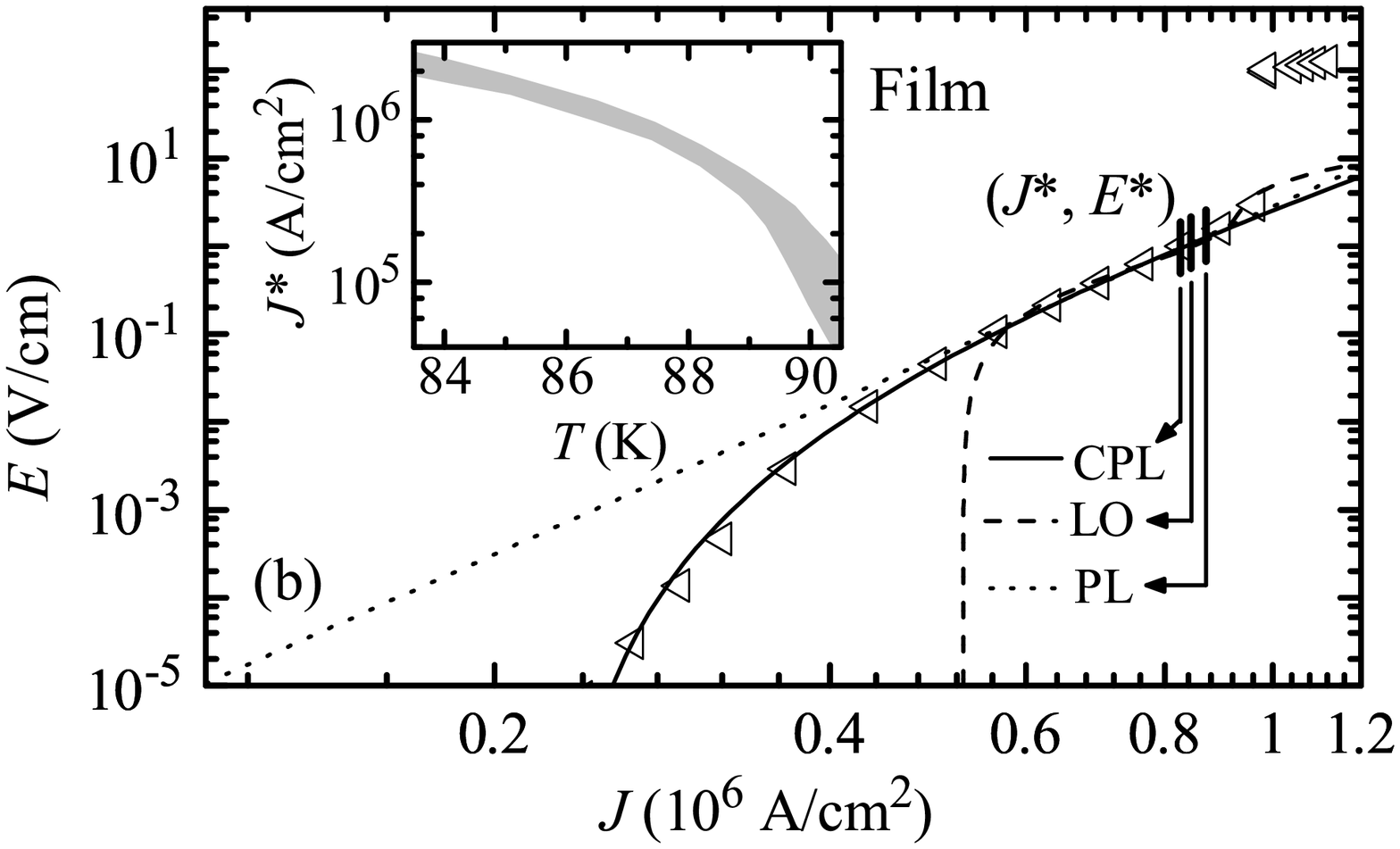}
\caption{\label{figFitOneT} Results of the fit of the three models studied to a
representative CVC, at $T/T_{c0}\simeq 0.97$, for: (a) melt-textured sample C11 and (b)
film Sys116. Dotted lines correspond to power law (PL) expression, dashed lines to LO
model and solid lines to critical power law (CPL). The $(J^*,E^*)$ coordinates,
determined by a threshold deviation criterion from the three theoretical expressions, are
marked over the $E$--$J$ curve. The insets summarize the bounds where the so obtained
$J^*$ for the CVCs at different temperatures are included in.}
\end{figure}}

\newcommand{\figFit}{
\begin{figure}
\includegraphics[width=7.3cm]{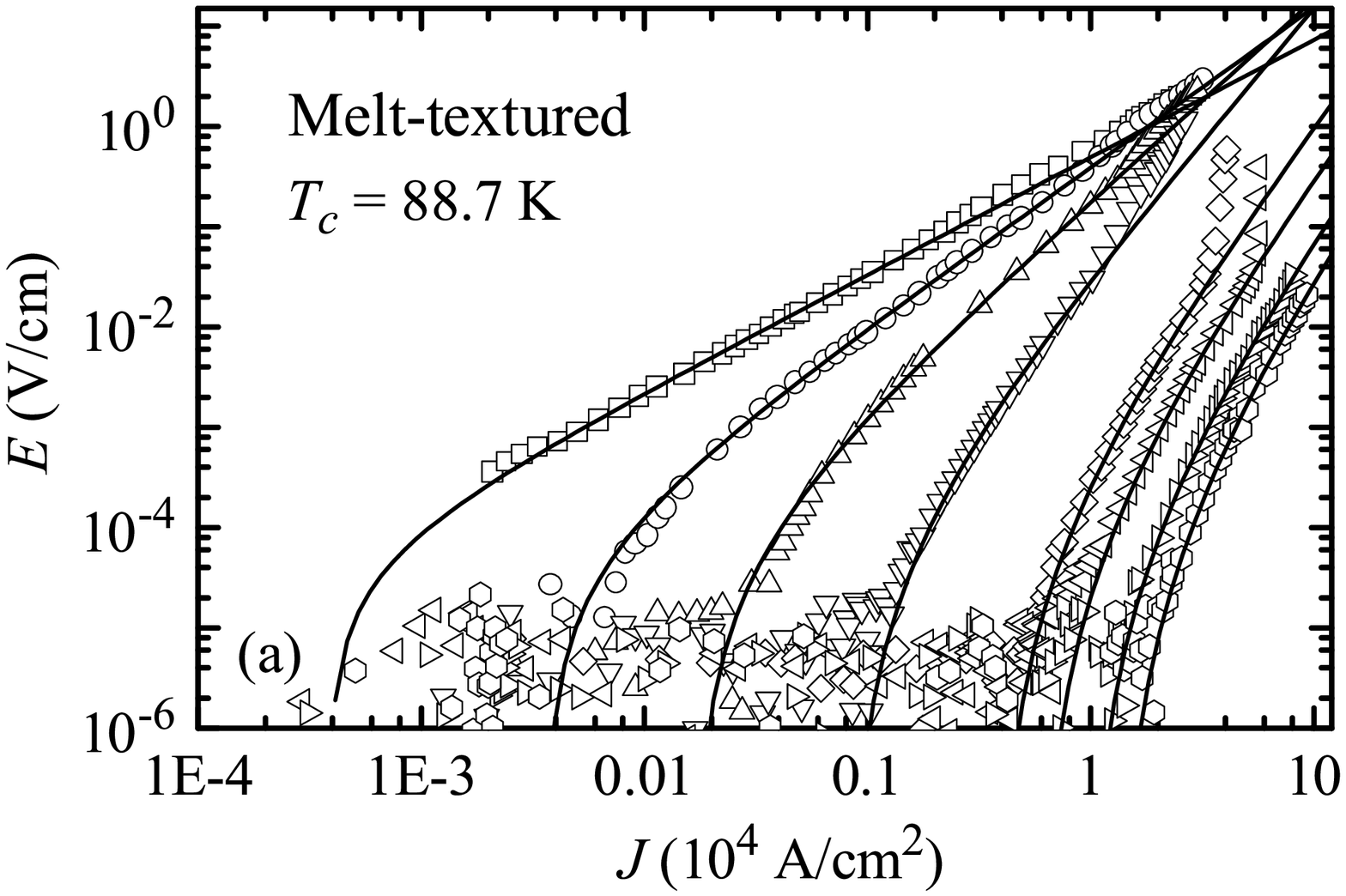}
\smallskip
\includegraphics[width=7.3cm]{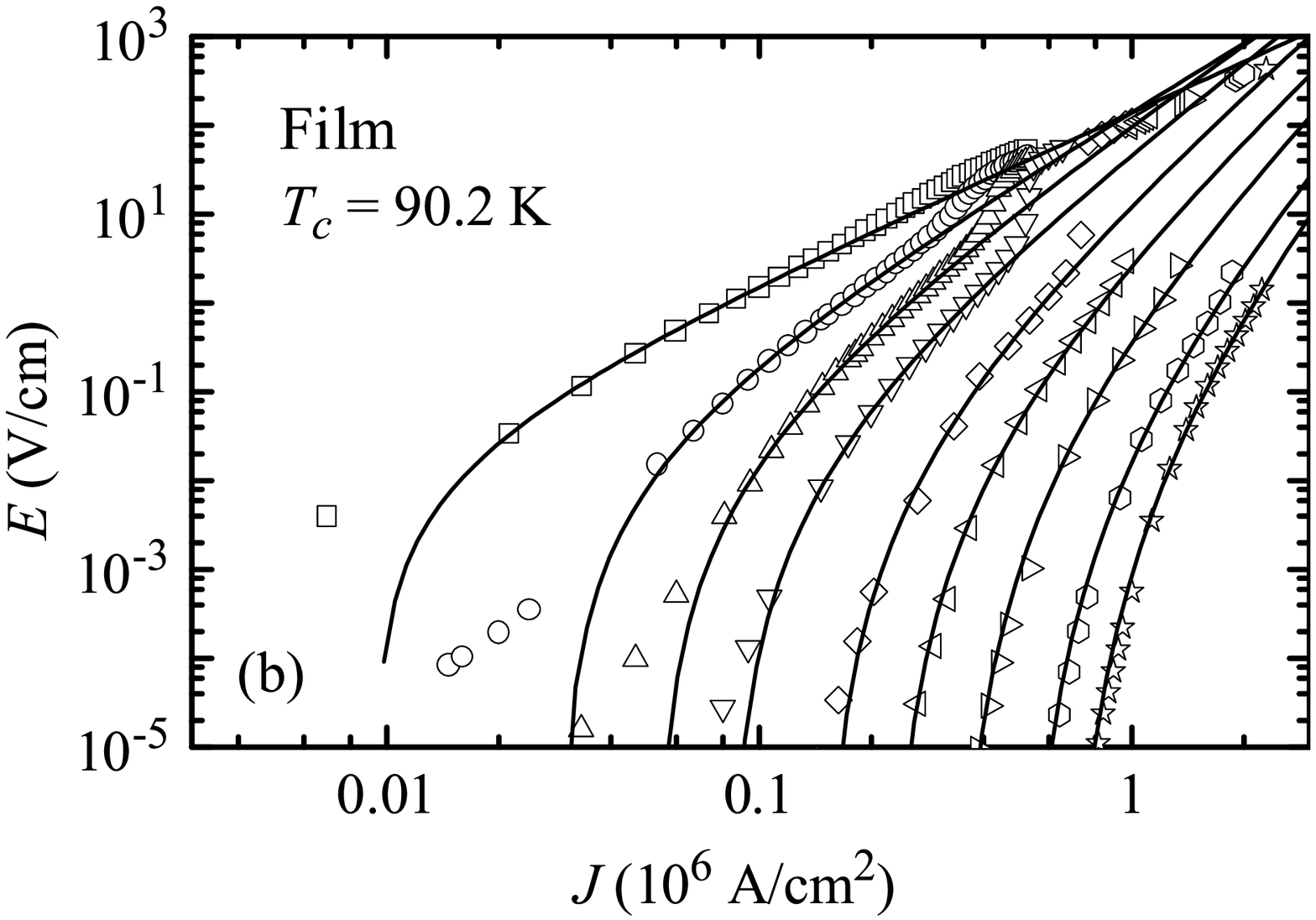}
\caption{\label{figEvsJFit} Results of the fit of the CPL expression given by
Eq.~(\ref{Vcritical}) to the $E$--$J$ data: (a) Melt-textured sample C11: from left to
right, the curves correspond to temperatures: 89.1, 88.5, 88.0, 87.5, 86.3, 85.7, 84.8,
and 83.9 K. (b) Film Sys116, with corresponding temperatures: 90.0, 89.7, 89.3, 88.9,
88.2, 87.4, 86.5, 85.1, and 84.1 K. The agreement is very good for all voltage ranges at
temperatures below the melting temperature.}
\end{figure}}

\newcommand{\figJcnvsT}{
\begin{figure}
\includegraphics[width=7.3cm]{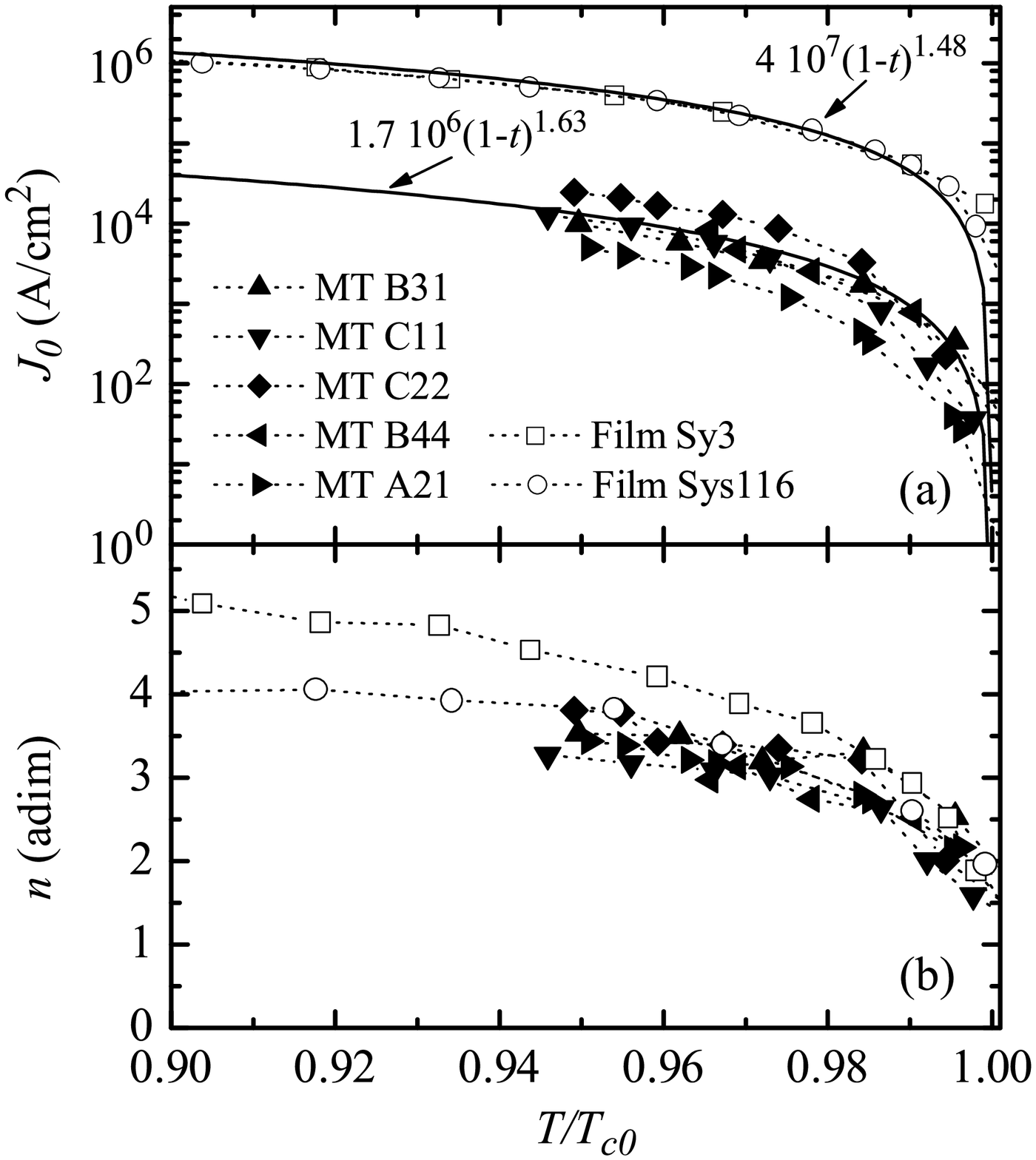}
\caption{\label{figJcnvsT} Variation with temperature of the parameters $J_{c0}$ and $n$
corresponding to the fit of the CPL expression to the CVCs for melt-textured samples and
films. $J_0$ has almost the same dependence on temperature for both kind of samples while
the values of $n$ are very similar at all temperatures studied. }
\end{figure}}

\newcommand{\figJqJcvsT}{
\begin{figure}
\includegraphics[width=7.3cm]{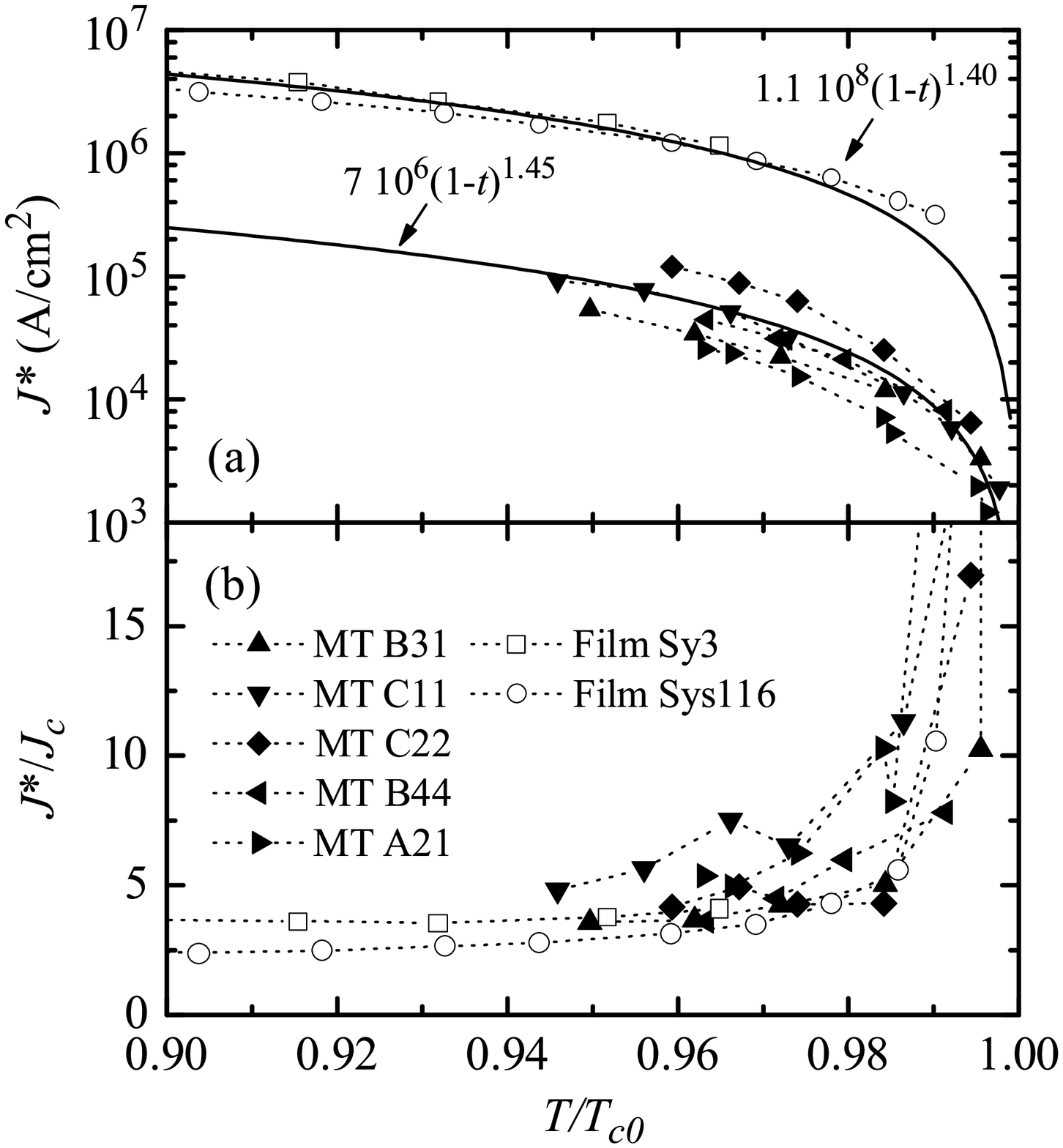}
\caption{\label{figJqJcvsT} (a) Variation of the quenching current density $J^*$ with
temperature. $J^*$ is determined by a threshold criterion of deviation from the
theoretical expression given by the critical power law. The temperature dependence of
$J^*$ for melt-textured samples is very close to that for films. (b) Correlation of $J^*$
and $J_c$ at different temperatures. Very similar ratio between these quantities is found
in textured monoliths and films. Dotted lines are only a guide for the eye.}
\end{figure}}

\newcommand{\grad}{\hspace{.1ex}\lower-1ex\hbox{\tiny{o}}}

\begin{document}


\title{Normal-superconducting transition induced by high current
densities in YBa$_2$Cu$_3$O$_{7-\delta}$ melt-textured  samples and thin films:
Similarities and differences.}

\author{M.~T.~Gonz\'alez}
\email{fmteresa@usc.es}
\author{J.~Vi\~na}
\author{S.~R.~Curr\'as}
\author{J.~A.~Veira}
\author{J.~Maza}
\author{F.~Vidal}
\affiliation{LBTS, Departamento de F\'{\i}sica da Materia Condensada\\ Universidade de
Santiago de Compostela E15782, Spain} \altaffiliation{Laboratorio de Bajas Temperaturas y
Superconductividad. Unit associated to the Instituto de Ciencias de Materiales de Madrid,
CSIC.}

\date{\today}

\begin{abstract}
Current-voltage characteristics of top seeded melt-textured YBa$_2$Cu$_3$O$_{7-\delta}$
are presented. The samples were cut out of centimetric monoliths. Films characteristics
were also measured on microbridges patterned on thin films grown by dc sputtering. For
both types of samples, a quasi-discontinuity or quenching was observed for a current
density $J^*$ several times the critical current density $J_c$. Though films and bulks
much differ in their magnitude of both $J_c$ and $J^*$, a proposal is made as to a common
intrinsic origin of the quenching phenomenon. The unique temperature dependence observed
for the ratio $J^*/J_c$, as well as the explanation of the pre-quenching regime in terms
of a single dissipation model lend support to our proposal.
\end{abstract}

\pacs{74.25.Fy, 74.25.Sv, 74.81.Bd, 74.72.Bk}

\maketitle

\section{Introduction}

Measurement of Current-Voltage Characteristics (CVCs) of high temperature superconductors
provides important information on critical current and dissipation mechanism, enhancing
our knowledge about the vortex dynamics of these materials  and about their practical
applicability.\cite{Blatter94, Tinkham} The mentioned properties are very sensitive to
microstructure (presence of grain boundaries, microcracks, twin-planes or different kinds
of bulk and surface defects), as well as to external conditions (temperature and magnetic
field). As a consequence, CVCs are expected to highly depend on these features as well.
In the low dissipative regime, the CVCs have been widely studied for polycrystalline
bulks, melt textured samples, films and single crystals. Several theories have been given
to explain the observed CVCs: flux flow, flux creep, collective flux pinning, vortex
glass transition...\cite{Blatter94, Tinkham}

The highly dissipative regime at high current densities has received attention recently
and results are available for films.\cite{Samoilov95,
Klein85,Lefloch99,Curras01,Xiao96,Xiao98b,Doettinger94,Doettinger95, Antognazza99,
Decroux00,Chiaverini01, Xiao98, Xiao99, Reymond02} The common observation is that, below
$T_c$, a transition to a highly dissipative regime takes place abruptly at a certain
current density $J^*$ several times the critical current density $J_c$. In this high
dissipative regime, the effective resistivity $\rho=E/J$ almost agrees with the normal
resistivity extrapolated to the sample temperature, $T<T_c$. Here, $E$ is the electric
field parallel to the applied current density $J$. This quasi-discontinuity in the CVCs,
sometimes called quench, has been observed for films of low critical temperature $T_c$
\cite{Samoilov95, Klein85,Lefloch99} and high critical temperature
(YBa$_2$Cu$_3$O$_{7-\delta}$ \cite{Curras01,Xiao96,Xiao98b,Doettinger94,Doettinger95,
Antognazza99, Decroux00} and Bi$_2$Sr$_2$CaCu$_2$O$_{8+\delta}$ \cite{Chiaverini01,
Xiao98, Xiao99, Reymond02}) superconductors. The quench is usually explained by the
Larkin-Ovchinnikov theory, \cite{Larkin76} in terms of a high acceleration of vortices
when they reach a certain critical velocity, and by its modified version by Bezuglyj and
Shklovskij \cite{Bezuglyj} including heating effects as a refinement to the LO theory.
These theories have dropped the more classical explanation based in the existence of a
thermal bistability in these materials \cite{Skocpol74,Gurevich87,Skokov93}, thus only
needing heating effects to justify the existence of a quasi-discontinuity in the CVCs.
There are also other authors which explain the quench using arguments similar to the one
dimensional phase-slip centers theory.\cite{Jelila98,Maneval01,Reymond02}

Electrical transport measurements up to high current densities, inducing the transition
to the normal state, have been also carried out in melt
textured\cite{Porcar98,Tixador00,Tournier00,Yang99,Morita99,Morita00} (and melt
cast\cite{Herrmann97,Elschner99}) samples, only at 77~K or some other particular
temperature\cite{Tixador99}. Bulk superconductors are specially attractive because of
their great potential for applications at high transport currents. Again, a similar
abrupt increase of voltage has been observed in the CVCs of these materials. However, in
this case, the usual explanations do not come from intrinsic vortex dynamics, but from
thermal processes (thermal runaway \cite{Elschner99,Tournier00} or changes in the
dissipation regime of the samples\cite{Yang99}).

As a support for such divergent approaches to monoliths and films, phenomenological,
structural and morphological reason can be invoked. The more obvious difference between
the two kinds of samples is their critical current density, $J_c$. Melt-textured samples
have a $J_c$ around $10^4$--$10^5$ A/cm$^2$, for a typical section between 0.5 and 15
mm$^2$, while $J_c$ of films exceeds $10^6$ A/cm$^2$, for samples of typical thickness of
0.1-0.5 $\mu$m, and a more variable width, between 10 $\mu$m and 1 mm. Usually, this
difference in the $J_c$ values is attributed to structural causes: presence of more and
stronger pinning centers, lesser effect of possible grain boundaries or
microcracks\ldots, or also to current concentration on the outer surface of
samples\cite{Lazard02}. No less meaningful differences exist in their thermal
environment. The levity of films together with a close lattice matching with the
substrate favor a good refrigeration and a fast reaching of the steady state. Bulk
material, in turn, is massive and quite inevitably inhomogeneous, both features leading
to a poor refrigeration and hot spot development.

Despite these marked differences, the similarity of the overall quenching phenomenon seen
in the CVCs for films and bulk samples demands a more exhaustive measurement of the
latter, and closer comparative study between them. In fact, though there are
contributions that study both type of samples at the onset of dissipation at different
temperatures and magnetic
fields\cite{Gao91,Scurlock92,Solovjov94b,Brand94,Hettinger89,Gupta93,Cao97,Wang00}, we
are not aware of similar studies for melt-textured samples extending to high current
densities, and, in particular, the quenching effect. This paper is devoted to a
systematic comparison of dissipation and particularly the quench phenomenon in both
melt-textured (bulk) and thin films. The aim is to look closely into the phenomenological
behavior of charge transfer at high density in both systems, in order to infer a possible
shared origin of the quenching. This voltage jump is also very interesting for
applications of both melt-textured samples and films like fault current limiters, as it
means a fast change in the resistance of the device for currents above a critical value.

Systematic measurement of CVCs on melt-textured samples of YBa$_2$Cu$_3$O$_{7-\delta}$
(YBCO) have been carried out at various temperatures. For spurious heating to keep
moderately low, a near $T_c$ temperature range has been scanned, where critical current
values are lower. At the same time, refrigeration conditions were optimized by using a
pressurized cryostat so that samples were always submerged in liquid nitrogen at any
temperature. A comparison with CVCs measured in YBCO films at different temperatures was
carried out, studying the low dissipative regime before the quasi-discontinuity, as well
as the occurrence of the quasi-discontinuity itself.

\section{Experimental details}

\subsection{\label{MTexpdetails} Melt-textured samples}

Single domain bulk samples of YBa$_2$Cu$_3$O$_{7-\delta}$ plus 30 molar~\% Y$_2$BaCu$O_5$
and 0.1 wt.~\% Pt were prepared by top seeded melt growth (TSMG)\cite{Cardwell98,Lo96b}.
Commercial powders of YBa$_2$Cu$_3$O$_{7-\delta}$, Y$_2$BaCuO$_5$ and Pt were mixed in
the above proportions, and 30 g samples of this mixture were die pressed into 3 cm
diameter pellets. These were placed on MgO single crystal substrates\cite{Endo96} in a
box furnace, and heated to 1025\grad C to be decomposed peritectically. After a fast
decrease of temperature to 990\grad C, the samples were slowly cooled to around 960\grad
C at 0.3-0.4\grad C/h. In order to control the growth of YBCO during the cooling and
avoid the formation of a granular microstructure, a small seed of Nd-Ba-Cu-O was placed
over the samples. After the process, single domain monoliths up to 2 cm in diameter
(ab-planes direction) and 0.7 cm in height (c-axis direction) were grown. The
microstructure of the samples was studied using optical microscopy and their texture was
analyzed by x-ray diffraction. The general characteristics of these melt-textured samples
were much the same as those previously obtained following a similar procedure by Cardwell
and coworkers.\cite{Cardwell98,Lo96b}

Bars with typical dimensions of 0.5 mm$^2$ cross section and 7 mm in length were cut from
the monoliths, using a wire saw. We took the crystallographic ab-planes parallel to the
longest dimension, so that the current will flow along the ab-planes in the transport
measurements. Measurements of the variation of resistivity with temperature for these
samples were routinely made.

\figVt

As we have already mentioned, the critical current density in the ab-planes direction for
melt-textured samples is of the order of $10^4$ A/cm$^2$, which means current of hundreds
of Amperes for our samples. Moreover, we are interested in measuring the $V$--$I$ curve
for currents well above $I_c$. The experimental setup for transport measurements must
take these features into account, and be designed to minimize the unavoidable heating
effects. To reach this aim, the electrical contacts with the sample are one of the main
points. The samples were painted in their ends with silver paste and then annealed
briefly at 900\grad C, to favor the diffusivity of the silver into
them\cite{Parish91,Lepropre93,Svistunov98}. Then, they were cooled down to 500\grad C and
held at this temperature for several hours to finishing the contacts treatment and to
ensuring oxygenation\cite{Ekin93}. We obtain a typical contact resistance of 100
$\mu\Omega$ at the working range of temperature. The sample is connected with the rest of
the circuit through copper bars which are pressed to the sample by springs. An Indium
sheet was placed in between (avoiding the Indium to touch the sample\cite{Ekin93}) to
reduce the contact resistance.\cite{Maehata99} In this way, possible damages by
mechanical stress during the transport measurement are also prevented. A scheme of this
experimental arrangement is shown in Fig.~\ref{figPulseShape}(a).

The second factor to minimize heating effects is refrigeration. All our measurements were
made with the sample, plus a reference resistor in-series with it, placed in a hermetic
cavity filled up with liquid nitrogen. A valve lets control the pressure of the nitrogen
vapor in the cavity, and therefore control the temperature of the liquid nitrogen bath.
Furthermore, there is a nitrogen gas reservoir connected with the cavity for a better
control and stabilization of pressure.\cite{Diaz97,Curras01}

\figInduct

Finally, we minimized the time that the current goes through the sample. We worked with
1--3 ms long ramp shaped current pulses, as in certain cases longer pulses may burn out
the samples. For current application, we use a HP6681A power supply, which has fixed
raising ramps to a stable current of approximately 20 ms. Therefore, we used high speed
solid-state relays in order to cut the current and isolated 1--3 ms long ramped pulses.
Depending on the range of current and the resistance of the sample, the raising rate of
the supply varies between 5 and 35 kA/s. As we have limited the total length of the
pulse, the total $V$--$I$ curve were sometimes obtained as the result of several
different pulses. Figure~\ref{figPulseShape}(b) shows the shape of some of the current
pulses used in this work. During the current pulses, the voltage in the sample and in the
reference resistor  were recorded with a high speed data acquisition card (DAQ) of
National Instruments (PCI-4452) allowing the recording of several voltages
simultaneously. From the reference resistor voltage, the current flowing through the
circuit at any time was determined. By pressuring the hermetic cavity, we set the sample
immersed in liquid nitrogen at the critical temperature (around 90 K). $V$--$I$ curves
were measured for temperatures from this value down to temperatures where a significant
part of the curve is accesible within the limitations of our power supply.

After measuring the $V$--$I$ curves, the inductive component of voltage was subtracted:
Apart from the self-inductance, there is an inductive component due to the loop that each
element forms with the potential wires.\cite{Porcar97a} For the reference resistor, this
component is noticeable because its own resistance is very small. We deduced its
inductance from the deviation of the $V$--$I$ curve from the ohmic behavior when a high
resistive element is placed instead of the sample. For the sample, well below $T_c$, we
expect zero resistance for very low current values, so that all the observed voltage
should be inductive. Indeed, we found the sample voltage proportional to d$I$/d$t$ at
low, i.e. subcritical, current. An example of this last correction is shown is
Fig.~\ref{figEindvsJ}. We checked, at different current pulses, the independence of
inductance on current raising rate. Inductance is also found temperature independent over
our working range of temperature.

\subsection{Thin films}

\figSpets

\tabcharac

The two c-axis oriented YBCO thin films used for comparison were grown on (100)SrTiO$_3$
substrates by high pressure dc sputtering. Details about the growth technique and
characterization of the films are described elsewhere.\cite{Curras01} Microbridges of 10
$\mu$m wide were patterned by chemical wet photolithography and Au contact pads were
sputtered on them. After annealing, the typical contact resistance was 100 m$\Omega$. Al
wires were solded to these pads using a Kulicke \& Soffa 4523 wire bonder. CVCs at
different temperatures were measured in four probe configuration, using a different
procedure for each of the two films. The film named Sy3 was submerged in a liquid
nitrogen container, whose temperature was controlled through its pressure, as explained
previously for melt-textured samples. The CVCs were obtained using dc current supplied by
a HP6038A power source and cut by a relay: one voltage measurement was taken for every 30
ms constant current pulse, using a HP3457A multimeter.\cite{Curras01} On the other hand,
Sys116 film was set in He atmosphere and its temperature was regulated with an Oxford
ITC4 temperature controller. Current was supplied by a Keithley 2400 source, allowing the
application of dc current, or stepped ramps of about 1 ms of step length. For the lowest
dissipation regime, dc current was used in combination with a HP34420A nanovoltimeter.
For higher dissipation, stepped current ramps were measured with a National Instruments
PCI6035 DAQ, allowing us to discard inductive effects while acquiring the full CVC in a
single measurement of typically 20 ms. An example of these stepped ramps is shown in
Fig.~\ref{figSteps}. The results obtained by the ramps do not significantly differ from
those using isolated pulses, as samples seem to reach an almost steady state is less than
one millisecond. In fact, voltage was checked to be stable for square pulses around 30~ms
long, at currents at least 98\% the quenching current, in agreement with previous works
\cite{Curras01}. Only for currents very close to the quasi-discontinuity, a progressive
increase of the voltage signal was observed, probably due to heating of the sample. In
any case, each point in the CVCs is the mean value of voltage data read along a current
plateau.

\section{Results and discussion}

\subsection{\label{CVC} Current Voltage Characteristics}

Geometric parameters of our samples and the main results of their corresponding
$\rho$--$T$ measurements are summarized in Table~\ref{SamCharTable}. $T_{c0}$ is the
offset critical temperature, and $\rho_n$ is the extrapolation of the normal-state
resistivity to the working range of temperatures.

A typical set of CVCs at different temperatures for our melt-textured samples is plotted
in Fig.~\ref{figEJTText}, using the standard definitions: $J=I/A$ and $E=V/l$, where $A$
is the cross section area and $l$ the distance between voltage pads.
Fig.~\ref{figEJTText}(b) and (c) are closer views of the low voltage region of
Fig.~\ref{figEJTText}(a). Three current regions can be distinguished in the CVCs of
Fig.~\ref{figEJTText}. A first region of zero resistance, illustrated in
Fig.~\ref{figEJTText}(c). Then, a dissipative region where the voltage grows in a
non-linear but smooth way. Finally, there is a sharp change in the voltage up to a nearly
normal state behavior. The transition from one of these regions to the next can be
characterized by a given value of the current density: $J_c$ is the current density where
the dissipation begins, usually obtained by a threshold criterion in the electric field,
$E_c$, while $J^*$ is the current density where the quench in the CVCs occurs. As shown
in Fig.~\ref{figEJTText}(a), after the quench, the heat generated in the sample is too
high to be evacuated and the CVCs are diverted from the ohmic behavior. For temperatures
lower than 3 or 4 degrees below $T_{c0}$, the rapid increase in the sample resistance is
accompanied by a decrease of current (see the last pulse of Fig.~\ref{figPulseShape}(b)
for an example), as our power supply is not able to respond fast enough to the change of
charge. For this reason, at these temperatures, it was not possible to complete the CVC
up to the nearly normal state region.

As mentioned in Section~\ref{MTexpdetails}, we have used different pulses to measure the
CVCs of our melt-textured samples. We have observed a displacement of the quench current
to higher values as the initial current of the pulses applied gets higher. This result
reveals the importance of thermal effects in the occurrence of the quench. Nevertheless,
for temperatures lower than 2 degrees below $T_{c0}$, we have estimated a variation of
$J^*$ in less than 30\%, due to this effect, being smaller at lower temperature. This
dependence of the CVC results on the pulse features was observed only for currents near
the quench, whereas the CVC data for lower currents were well reproduced in different
pulses.

\figEJTte

\figEJTfi

As mentioned in the introduction, similar results have been previously reported for melt
textured samples at 77~K. As well, the overall behavior of these CVCs is common to films
or melt cast processed samples, but the explanation about the observed abrupt change in
the sample resistance at high current densities is variant. While in melt textured and
melt cast processed samples the quench is generally attributed to thermal runaway
processes,\cite{Tournier00,Yang99,Elschner99} there are several authors which explain the
abrupt resistance change observed in films as a high flux-flow velocity
instability.\cite{Doettinger94,Xiao96,Chiaverini01} Here, we will closely compare the
results obtained for melt-textured samples with similar measurements of CVCs in YBCO thin
films at different temperatures. Figure~\ref{figEJTFilm} shows the corresponding set of
CVCs for the film Sys116. Comparison with Fig.~\ref{figEJTText} confirms that indeed the
general behavior of films matches that of melt-textured samples. We can clearly
distinguish three different dissipative regions: zero resistance, non-linear $V$--$I$
curve above a certain critical current, and near ohmic behavior after an abrupt voltage
discontinuity. On the other hand, this comparison also permit to pointed out a main
difference between the CVCs of melt-textured samples and films. As may be checked in
Figs.~\ref{figEJTText} and~\ref{figEJTFilm}, both $J$ and $E$ along the whole CVCs are
approximately two orders of magnitude higher in films, and this includes $J^*$ and $E^*$.
The abundant results on transport properties for currents near the beginning of
dissipation both for melt-textured samples\cite{Gao91,Scurlock92,Solovjov94b,Brand94} and
films\cite{Hettinger89,Gupta93,Cao97,Wang00} had already well established the existence
of approximately a factor $10^2$ between $J_c$ values of both kind of samples. From the
observation of the CVCs presented in this work, it can be stated that this feature is
maintained along the whole CVCs. As a result, the CVCs in both type of samples look
scalable in that they differ both in $J$ and $E$ in a constant factor.

\figJcvsT

Regarding $J_c$, Fig.~\ref{figJcvsT} illustrates the comparison between its values for
melt-textured samples and films. Despite the different magnitude, $J_c$ shows a
dependence on temperature close to $(1-t)^{3/2}$, where $t=T/T_{c0}$, for both
melt-textured samples and films. In this figure, $J_c$ has been determined by the
threshold criterion of 10 $\mu$V/cm. The estimation of $J^*$ and $E^*$ is a bit more
complicated. Looking at the curves of Fig.~\ref{figEJTFilm}(a) and (b) at low
temperature, the quenching point seems easy to be obtained because of the neat jump in
voltage. However, the corresponding voltage variation is much less abrupt at higher
temperature, close to $T_c$. Furthermore, bulks show smoother quenching than films (see
Fig.~\ref{figEJTText}). In all, a systematic criterion for ($J^*$, $E^*$) determination
is needed. By observation of CVCs for several samples and temperatures, we found no
specific value of the electric field, power or resistivity, nor a value of their first
derivatives, where the quasi-discontinuity systematically occurs. For this reason, a good
option is to use a deviation criterion from a background functional form, the latter
taken from the agreed-on models for CVCs.

In the next section we study the agreement of several dissipation models to our CVC
measurements. Apart from providing us with a functional background for the later
extraction of $(J^*, E^*)$, this correlation will permit us to obtain a detailed
comparison between the dissipation mechanism in melt-textured samples and films.

\subsection{\label{Models} Dissipation models}

\figFitOneT

The probably most popular model to describe non-linear CVCs of type-II superconductors is
the thermally activated flux creep theory \cite{Blatter94, Tinkham}. According to this
model, the first observation of non-zero voltage occurs below the depinning critical
current $I_c^{pin}$, due to the displacement of vortices from a pinning center to other
caused by thermal fluctuations. The generalized expression for the voltage at a given
current below $I_c^{pin}$ is:
\begin{equation} \label{FluxCreepV}
V=V_c^{pin} e^{-\frac{U_0(T)}{k T \mu} \left[\left(I/I_c^{pin}\right)^{-\mu}-1\right]}
\end{equation}
This generic expression comprises different previously proposed models depending on
$\mu$: $\mu=-1$ for the classical Anderson-Kim flux creep model, valid for conventional
type-II superconductors, where the pinning is strong, and the dissipation starts for
current near $I_c^{pin}$. $\mu= 7/9,3/2,1/7$(depending on the current range, magnetic
field and temperature) for collective creep/vortex glass models, which assume  the
beginning of a collective movement of vortices at $I\ll I_c^{pin}$ in materials with weak
pinning, as it is the case of high temperature superconductors. Finally, $\mu\rightarrow
0$, for the logarithmic Zeldov model, which is a good approximation of the latter for the
high current regime ($I\alt I_c^{pin}$) and leads to the simple expression:
$V=V_c^{pin}(I/I_c^{pin})^n$, with $n=U_0/(k T)$.

The generalized expression has the inconvenience of having too many free parameters. To
begin with, a given value of $\mu$ is only expected to describe a limited region of
current. Furthermore, presetting the value of $I_c^{pin}$ or $V_c^{pin}$ is needed,
because otherwise multiple valid solutions for voltage, $V$, are obtained. Their values
should be the coordinates of the transition point from a non-linear $V$-$I$ curve
(flux-creep regime) to a linear one (flux-flow regime). However, such a transition is not
observed in our results, i.e. the CVCs are always non-linear up to the voltage
quasi-discontinuity. Some authors \cite{Xiao97} claim that the quasi-discontinuity in the
CVCs may occur at $I_c^{pin}$ and then $I_c^{pin}=I^*$ and $V_c^{pin}=V^*$. Even if this
is the case, we can not use this information as an input, because $(I^*,V^*)$ is an aside
information that we want to extract from the analysis. For these reasons, we used the
power law (PL) expression which can be written as:
\begin{equation} \label{VZeldov}
V=A_c I^n,
\end{equation}
independently on the particular $(I_c^{pin}, V_c^{pin})$ election.
Figure~\ref{figFitOneT} shows an example of the fit of the model (dotted lines) to the
CVCs for both bulks and films, at a given temperature. The model fits satisfactorily the
high current region of the CVCs before the quench, which is the range we are mainly
interested in.

Another model suggested in order to explain non-linear $I$-$V$ curves, and linked to the
nature of the quench, is the non-linear flux-flow theory by Larkin and Ovchinnikov
\cite{Larkin76} (LO). According to this model, the viscosity opposing the movement of
vortices in the flux-flow regime varies with their velocity, leading to a non-linear CVC.
Moreover, there is a certain critical velocity where viscosity reaches its maximum and
then decreases. This causes a rapid acceleration of vortices, which would lead to an
abrupt increase of voltage. As already mentioned, several authors state that this could
be the correct explanation for the quench observed in
films\cite{Doettinger94,Xiao96,Chiaverini01}. Below the predicted quasi-discontinuity,
the expression linking $V$ and $I$ is given by:
\begin{equation} \label{VLO}
\left[ \frac{V}{1+(V/V^*)^2}+V \left( 1 - \frac{T}{T_c} \right)^{1/2}\right]
=R_f(I-I_c^{pin}),
\end{equation}
where $R_f$ is the resistance in the flux-flow regime. This expression is claimed to fit
adequately measurements by other groups\cite{Klein85,Doettinger94}. However, as shown in
Fig.~\ref{figFitOneT} (dashed lines), it only fits our measured CVCs very near the
transition, both for bulks and films. Moreover, the agreement with our data is not better
than the one obtained with the power law.

\figFit

Finally, we have also inspected another proposed model for non-linear CVCs, which we term
critical power law (CPL):
\begin{equation} \label{Vcritical}
V=V_{c0}(I/I_0-1)^n,
\end{equation}
where $I_0$ is the critical current which dissipation begins at. This expression, as
happens with the power law given by Eq.~(\ref{VZeldov}), may be derived from various
physical backgrounds.\cite{Prester98} In reference to the vortex dynamics, this
expression can be obtained modifying the Zeldov expression for thermally activated flux
creep model by introducing a threshold to the activation of vortices movement by thermal
fluctuations.\cite{Antognazza02} It can also be deduced in the mean-field approximation
for the case of strong pinning and neglecting the effect of thermal
fluctuations.\cite{Solovjov94b} On the other hand, the same voltage dependence on current
is found in the context of granular materials modelized as an array of Josephson
junctions.\cite{Prester98}

\figJcnvsT

A good agreement has been previously found of the critical power law with CVCs of single
crystal and melt-textured samples\cite{Solovjov94b}, and films\cite{Antognazza02}, only
at limited temperature or current ranges. In this work, we have found a satisfactory fit
of this expression to all our CVCs with a value of $V_{c0}$ independent on temperature,
leaving two free parameters, $I_0$ and $n$, for each $I$--$V$ curve. The improvement in
the fit respect to PL or LO expressions, specially for films, is clearly shown in
Fig.~\ref{figFitOneT}. The agreement with our data is indeed very good over all voltage
ranges, as illustrated in Fig.~\ref{figEvsJFit}. Only for high temperatures, close to
$T_c$, the fit is not good at low current. These CVCs feature a thermally activated
flux-flow regime (TAFF) which indicates that we are above the melting
temperature,\cite{Blatter94} and, hence, neither Eq.~(\ref{FluxCreepV}) nor
Eq.~(\ref{Vcritical}) are expected to hold.

Note that the good agreement of the critical power law with CVCs measured in absence of
an external magnetic field should not be surprising: Flux creep theory in materials with
weak pinning ($0\leq \mu \leq 1$) states that, even at very low current, thermal
fluctuations can cause displacements of vortices from their pinning centers, thus
generating a voltage. However, this is the case only in experiments in which vortices are
previously created by an external magnetic field above $H_{c1}$. When no external field
is applied, as in our data, the sample will stay in the Meissner state up to currents
producing a self-field higher than $H_{c1}$. For lower currents, no vortices are still
present and there is no dissipation. Later, for higher currents, vortices start to be
created and may begin to creep. As a result, a threshold for the beginning of
dissipation, as the CPL expression features, must be necessary to explain the CVCs.

In order to compare the results for melt-textured samples and films, we have plotted in
Fig.~\ref{figJcnvsT} the values of the parameters obtained from the fit of the CPL
expression, for both geometries. The parallelism found in the behavior of these
parameters suggests a common dissipation mechanism in melt-textured samples and films. In
particular, note that the values of $n$ are very close in both kinds of samples, while
the values of $J_0=I_0/A$, differing approximately in two orders of magnitude, have
almost the same dependence on temperature. Furthermore, the values obtained for
$E_{c0}=V_{c0}/l$ are very similar for samples of the same kind. We have $E_{c0}=10^{-2}$
V/cm, for films, and $E_{c0}= 5\;10^{-5}$ V/cm, for melt-textured samples, reflecting the
more than two orders of magnitude difference in $E$ in the CVCs themselves.

To summarize, a single model describing the existence of a threshold for the beginning of
dissipation accounts for the CVCs measured on both massive and film samples. With the
exception of high temperature, where the critical behavior itself blurs, all current and
temperature ranges studied are very well described by the CPL expression. An appealing
feature is that only three parameters, one of them temperature independent, suffice to
explain dissipation in our YBCO superconductors. The model's parameters follow a parallel
temperature behavior for monoliths and films, thus pointing to a unique underlying
dissipation mechanism. These results in turn would lend support to a common origin for
the starting of the quenching phenomenon. This point is given a closer look in the next
paragraph.

\subsection{The quenching phenomenon}

\figJqJcvsT

As mentioned in Section~\ref{CVC}, we determined $(J^*, E^*)$ from a deviation criterion
from the background dissipation. Since in the high voltage region three dissipation
models fit the data (see the previous section), the sensitivity of $(J^*, E^*)$ on the
choice of the background dissipation model may be tested.

We set the deviation threshold value to be above the noise of our data: 1 mV/cm for
melt-textured samples, and 10 mV/cm for films. The values of $(J^*, E^*)$ were obtained
in this way for the PL and CPL expressions. LO expression has the quench voltage as a
parameter, so that $(J^*, E^*)$ are obtained directly from the fit. As an example, the
resulting $(J^*, E^*)$ at $T/T_{c0}=0.97$ are marked in Fig.~\ref{figFitOneT}(a) and (b),
for the melt-textured sample C11 and the film Sys116, respectively. The so obtained $J^*$
at different temperatures are included in the bounds plotted in the Fig.~\ref{figFitOneT}
insets. Note that there is only a weak sensitivity of $(J^*, E^*)$ to the dissipation
model choice, except for temperatures very close to $T_c$, where the uncertainty is also
higher.

Figure~\ref{figJqJcvsT}(a) summarizes the resultant $J^*$ for our samples. As indicated
in this figure, $J^*$ for the melt-textured samples shows a dependence with temperature
very close to $(1-t)^{3/2}$, which is the same result found in films here, and in other
works \cite{Curras01,Xiao98b,Antognazza02}, for temperatures near $T_c$.
Figure~\ref{figJqJcvsT}(b) shows also that the ratio $J^*/J_c$ is very similar for both
kinds of samples at any temperature. For temperatures below 0.98 $T_{c0}$, $J^*$ tends to
a temperature independent value between 3 and 5 times $J_c$ (check also
Table~\ref{SamCharTable}). This result, already reported for films
\cite{Decroux00,Antognazza02}, is found here to be also common to melt-textured samples.

The unifying trend showed by the ratio $J^*/J_c$ would weaken the sometimes implicitly
accepted dual causes for the quenching, namely thermal, i.e., nonintrinsic, for massive
superconductors, and vortex dynamics or depinning limit, i.e., intrinsic, for films.
Indeed, so similar results suggest that the origin of the phenomenon is common for both
kind of samples, and this fact can give us some hints about its nature.

Thermal effects are indeed present on our CVCs. On general grounds, however, the very
different thermal conditions for films and monoliths work against the thermal origin of
the quenching. As mentioned in the introduction, the thermal interchange with the
surroundings is much better in thin films than in melt-textured samples. Due to the low
heat transfer coefficient of YBCO with liquid nitrogen (0.1-1 W/cm$^2$K
\cite{Tournier00,Mosqueira93b}), electric power dissipation for massive samples, at the
millisecond scale, takes place practically adiabatically. On the other hand, for films,
the heat transfer coefficient with the SrTiO$_3$ substrate is around 700-1000 W/cm$^2$K
\cite{Gupta93,Xiao98b}, favoring a fast heat diffusion. In fact, the low dependence of
$J^*$ on the total time of the pulse above around 1~ms,\cite{Doettinger94,Xiao99} and the
stability of $E$ at constant $J\alt J^*$ applied, during tens or hundreds of milliseconds
\cite{Curras01}, support that, in our experimental conditions, a quasi steady state is
reached in films, even very near $J^*$.

Note also that the dissipated power density $W=E\;J$ is, in average, nearly four orders
of magnitude higher in films than in monoliths, as easily computed from
Figs.~\ref{figEJTText} and ~\ref{figEJTFilm}. Furthermore, the results for thin films
have been proved to be quite independent of thermal conditions, as our two samples
studied have very similar $J^*$ and $J^*/J_c$ values in spite of the different
experimental environments and procedures (same result was also obtained by other
groups\cite{Xiao96}). In all, there seems to be too many relevant differences in their
thermal environment to easily ascribe the convergence of $J^*/J_c$ onto a single curve
shown in Fig.~\ref{figJqJcvsT}(b) to only thermal effects.

Notwithstanding, in both kind of samples (thin films and bulks) thermal effects indeed
influence our CVCs. For instance, we have already mentioned the sensitivity of the CVCs
of melt-textured samples to the current pulse characteristics for current near $J^*$,
giving a variability of a few tens per cent; certainly these differences are attributable
to thermal causes. It is likely that thermal effects together with other source of
extrinsic effects as sample inhomogeneity could justify the relatively high data
scattering in Fig.~\ref{figJqJcvsT} (more precisely, data scattering would be compatible
with heating effects contributing of the order of 50\% the total magnitude). Of course,
only a thorough  study on the thermal dynamics can put in more precise figures on the
thermal effects on the quenching. Anyhow, any convincing model given to explain the
quenching phenomenon should account for the similar results found in this work for YBCO
melt-textured samples and thin films, despite their different environment conditions.

\section{Conclusions}

Current-Voltage Characteristics (CVCs) of YBa$_2$Cu$_3$O$_{7-\delta}$ melt-textured
samples and films have been measured at several temperatures close to $T_c$. In order to
minimize heating effects, a pulse technique was used, and refrigeration conditions were
optimized by using a pressurized cryostat so that samples were always submerged in liquid
nitrogen at any temperature.

From the comparison of both sets of curves, we found an overall common behavior: at low
enough applied current a region of zero resistance is observed, followed by a non-linear
$I$--$V$ curve of low dissipation, and, finally, after an abrupt increase in the
resistance of the sample, a highly dissipative region of nearly ohmic behavior is
reached.

We have found that the CVCs for both kind of samples share the same features, apart from
approximately a factor $10^2$ both in $J$ and $E$ values. Thus, a simple theoretical
expression, namely a critical power law, has been found to reproduce the pre-quenching
dissipative region of the CVCs at different temperatures, with almost the same exponent
and amplitude's temperature dependence. The crossover current densities, $J_c$ (critical
current density) and $J^*$ (quench current density), are also found to follow the same
dependence on temperature. More importantly, in spite of the difference of almost two
orders of magnitude between their values in bulk samples and films, their ratio is
observed to collapse onto a single temperature dependent curve, common to both kinds of
samples. These findings point to an common origin of the CVC behavior in films and
melt-textured samples including the quenching, despite the usual different treatment in
the literature.

We have also described the very dissimilar role that thermal coupling, shape and thermal
inertia are to play on the current-induced heating of bulks and films, suggesting that
the final common cause of the quenching phenomenon is primarily intrinsic. The measured
fine-scale behavior of $J^*(T)$ can however be modulated by thermal effects.

Further evidence on this conclusion relies on working out some associated issues. In
particular, how to bound thermal effects on the observed CVCs? Likely, finite element
analysis would help ascertain whether heating accounts for the wide data scattering of
the ``universal'' ratio $J^*/J_c$. Secondly, despite the progress towards a common
descriptive frame for superconducting monoliths and films reported here, the
two-orders-of-magnitude difference in both current density and electric field amplitudes
is demanding an explanation. For instance, why are the $J_c$ or the $J^*$ values so
different but not their ratio? The study of these issues, including the influence of the
current distribution in the samples, likely to be dependent on the CVCs regimes, is
presently under way.

\begin{acknowledgments}

This work has been financed by the CICYT (MAT2001-3053 and MAT2001-3272), Uni\'on Fenosa
(contract no. 0666-2002) and Xunta de Galicia (PGIDIT01-PXI20609-PR, PGIDIT02-PXI20610-PN
and PGIDIT02-PXI20609-PN). MTG wishes to acknowledge Prof. D. A. Cardwell and his group
for their hospitality in the IRC in Superconductivity at Cambridge University, and their
invaluable help in getting familiar with melt-texture synthesis techniques.
\end{acknowledgments}

\end{document}